\documentclass[%
superscriptaddress,
 amsmath,amssymb, aps, prc,
floatfix, longbibliography, dvipsnames, twocolumn ]{revtex4-2}

\usepackage{graphicx}
\usepackage{dcolumn}
\usepackage{bm}
\usepackage{orcidlink}
\usepackage{graphics}
\usepackage{amsthm}
\usepackage{amsmath}               
  {
      \theoremstyle{plain}
      \newtheorem{assumption}{Assumption}
  }
\usepackage{quantikz}
\usepackage{makecell}
\usepackage[caption=false]{subfig}

\usepackage{glossaries}
\newacronym{hep}{HEP}{High Energy Physics}
\newacronym{annni}{ANNNI}{Axial Next-Nearest Neighbor Ising}
\newacronym{dmrg}{DMRG}{Density Matrix Renormalization Group}
\newacronym{qag}{QAG}{Quantum Angle Generator}
\newacronym{qcbm}{QCBM}{Quantum Circuit Born Machine}
\newacronym{qml}{QML}{Quantum Machine Learning}
\newacronym[
	first={Parametrized Quantum Circuit (PQC)},  
	firstplural={Parametrized Quantum Circuits (PQCs)}, 
	plural={PQCs}  
]{pqc}{PQC}{parametrized quantum circuit}  
\newacronym{nisq}{NISQ}{Noisy Intermediate-Scale Quantum}
\newacronym{vqe}{VQE}{Variational Quantum Eigensolver}
\newacronym{mmd}{MMD}{Maximum Mean Discrepancy}
\newacronym{mse}{MSE}{Mean Squared Error}
\glsdisablehyper

\theoremstyle{definition}

\theoremstyle{remark}

\begin{document}

\preprint{APS/123-QED}

\title{Symbolic Pauli Propagation for Gradient-Enabled Pre-Training of Quantum
	Circuits}
\author{Saverio Monaco \orcidlink{0000-0001-8784-5011}}
\affiliation{RWTH Aachen University, Aachen 52062, Germany}
\affiliation{Deutsches Elektronen-Synchrotron DESY, 22603 Hamburg, Germany}

\author{Jamal Slim \orcidlink{0000-0002-9418-8459}}
\affiliation{Deutsches Elektronen-Synchrotron DESY, 22603 Hamburg, Germany}

\author{Florian Rehm \orcidlink{0000-0002-8337-0239}}
\affiliation{European Organization for Nuclear Research (CERN), Geneva 1211, Switzerland}

\author{Dirk Kr\"ucker \orcidlink{0000-0003-1610-8844}}
\affiliation{Deutsches Elektronen-Synchrotron DESY, 22603 Hamburg, Germany}

\author{Kerstin Borras \orcidlink{0000-0003-1111-249X}}
\affiliation{RWTH Aachen University, Aachen 52062, Germany}
\affiliation{Deutsches Elektronen-Synchrotron DESY, 22603 Hamburg, Germany}

\date{\today}

\begin{abstract}
	Quantum Machine Learning models typically require expensive on-chip training
	procedures and often lack efficient gradient estimation methods. By
	employing Pauli propagation, it is possible to derive a symbolic
	representation of observables as analytic functions of a circuit's
	parameters. Although the number of terms in such functional representations
	grows rapidly with circuit depth, suitable choices of ansatz and controlled
	truncations on Pauli weights and trigonometric degrees yield accurate yet
	tractable estimators of the target observables. With the right ansatz
	design, this approach can be extended to system sizes beyond the reach of
	classical statevector simulation, enabling scalable training for larger
	quantum systems. This also enables a form of classical pre-training through
	gradient-based optimization prior to deployment on quantum hardware. The
	proposed approach is demonstrated on the Variational Quantum Eigensolver for
	obtaining the ground state of the ANNNI spin model on 32 qubits, showing
	that accurate results can be achieved with a scalable and computationally
	efficient procedure.
\end{abstract}

\maketitle


\section{Introduction}
\label{sec:intro}

\Glspl{pqc} have attracted considerable attention in recent years due to their
potential applications in \gls{qml} and the growing availability of hardware
capable of executing quantum algorithms for tasks of practical interest.

Despite the progress in hardware, a major challenge for \gls{qml} remains the
training time of these models. Evaluating and optimizing \glspl{pqc} on quantum
hardware is resource-intensive, with the added consideration that quantum
devices are currently limited and experimental.

Obtaining gradients for \gls{pqc} further exacerbates the problem. Computing
analytical gradients via the parameter-shift~\cite{paramshift} rule requires a
number of circuit evaluations that scales with the number of parameters. For
more realistic tasks, such as generating a single MNIST
image~\cite{deng2012mnist}, the estimated time to perform a single training step
can reach prohibitive times~\cite{trainonclassical}, making straightforward
scaling unfeasible. Gradient-free optimizers, such as SPSA~\cite{spsa},
COBYLA~\cite{cobyla}, and their variants, require a constant number of circuit
evaluations per step and offer a partial workaround. However, these methods are
often unstable in practice and do not fully alleviate the dependence on costly
quantum hardware for training.

To mitigate these challenges, recent efforts have focused on designing
gradient-efficient ansätze and formalizing tasks through the evaluation of
expectation values. In these approaches, the estimation of observables and their
gradients can be classically simulable~\cite{classicallyestimating} or
efficiently computed using a limited number of circuit
evaluations~\cite{backpropscaling}, while the output remains classically hard to
simulate. Examples include the \gls{vqe}, which minimizes the expectation value
of a Hamiltonian, and generative tasks whose goal can be formalized as matching
the expectation values of randomly sampled Pauli strings in the $Z$ basis to the
corresponding values from the training set~\cite{trainabilitymmd}.

In this work, we propose a method to represent an observable as an explicit
function of the circuit parameters in symbolic form, independent of any
particular parameter assignment.

This approximate representation can be used to pre-train \glspl{pqc} for the
tasks described above, which can then be fine-tuned via on-chip training or
alternative optimization techniques.

In general, computing the exact functional representation of a circuit's
observable is computationally challenging, as the number of terms grows (in the
worst case scenario) exponentially with the number of parameters. To address
this, we introduce two truncation schemes: a cutoff on the Pauli weight, which
is particularly effective for locally scrambling
ansätze~\cite{classicallyestimating}, and a cutoff on the trigonometric degree, defined as
the number of sine and cosine terms multiplied together in each product. These
cutoffs enable an efficient approximation of the observable while keeping the
computational complexity under control.

The paper is organized as follows. In Section~\ref{sec:methods}, we introduce
the framework for the \emph{propagation of observables} in quantum circuits. We
begin in Section~\ref{subsec:exact} by describing the mechanism of exact
propagation of observables in the Heisenberg picture.

Section~\ref{subsec:explicit} examines the exact representation of circuits
motivating the need for truncation schemes.

Sections~\ref{subsec:truncW} and \ref{subsec:truncV} provide the justification
for the truncation schemes, which enable efficient propagation of observables at
qubit numbers beyond classically simulable regimes.

Our results are presented in Section~\ref{sec:res}, where we apply the method
within the framework of the \gls{vqe} in Section~\ref{subsec:vqe}, using the
\gls{annni} model~\cite{annnioriginal} as a test case. We justify the use of
locally scrambling circuits for local spin models and compare the
propagated-energy estimates with exact \gls{dmrg} results.

Finally, in Section~\ref{sec:conc}, we discuss other classes of problems to
which this framework can be effectively applied and outline potential directions
for improving the Pauli propagation method.

\section{Propagation of Observables}
\label{sec:methods}
\subsection{Exact propagation}
\label{subsec:exact}

In quantum computation, the Schr\"odinger representation is often used because
it aligns intuitively with the circuit-based model, where the quantum state
evolves as gates are applied. Given an initial state $|\psi\rangle$, the
application of a unitary operator $U$ transforms the state as
\begin{equation}
	|\psi\rangle \to U |\psi\rangle.
\end{equation}

An alternative perspective is offered by the \textit{Heisenberg representation}
\cite{heisenberg}, in which the evolution of observables is tracked instead of
the quantum state. In this framework, the observable evolves \textit{backwards}
through the circuit via a similarity transformation,
\begin{equation}
	\label{eq:similarity}
	O \to U^\dagger O U.
\end{equation}
This backward evolution is referred to as \textit{Pauli propagation} when
restricted to the Pauli basis, since any observable can be expressed as a sum of
Pauli operators.

During propagation, non-Clifford gates~\cite{clifford} induce a
\textit{branching effect}, where a single Pauli word (a tensor product of Pauli
operators) evolves into a linear combination of two Pauli words. This branching
causes the number of terms to grow exponentially with circuit depth. For
instance, applying a rotation gate $R_Y(\theta)$ on the Pauli operator $X$
results in
\begin{equation}
	\label{eq:similarityRYX}
	R_Y^\dagger(\theta) X R_Y(\theta) = \cos(\theta) X + \sin(\theta) Z,
\end{equation}
where the original operator $X$ branches into a combination of $X$ and $Z$
weighted by trigonometric functions of the parameter $\theta$.

Unlike previous approaches such as Ref. \cite{classicallyestimating} and Ref.
\cite{dualrolelowweightpauli}, in this work the propagation is carried out
\emph{symbolically}, leaving the parameters $\vec{\theta}$ unassigned.
Consequently, the final propagated observable is an analytic function of the
circuit parameters.

Following the propagation through the similarity transformation in
Eq.~\ref{eq:similarity} and the matrix representation of the circuit's gates,
two other key effects can be observed. First, two-qubit gates typically increase
the \textit{Pauli weight} of the propagated words, that is, the number of
non-identity Pauli operators acting on different qubits. Second, each
parametrized gate associated with a parameter $\theta_i$ multiplies the
coefficient of the corresponding Pauli word by $\sin(\theta_i)$ or
$\cos(\theta_i)$, while simultaneously branching it into two distinct terms, as
seen in Eq. \ref{eq:similarityRYX}.

We define the \textit{trigonometric degree} of a Pauli word as the number of sine and
cosine factors it contains.

After the observable has been fully propagated through the circuit, it becomes a
sum of Pauli words with parameter-dependent coefficients. To obtain its
expectation value, we sandwich each Pauli word with the initial state
$\ket{0}^{\otimes n}$. During this step (referred to later as \emph{trimming}),
all Pauli words that contain any $X$ or $Y$ operators vanish, since their
expectation value on $\ket{0}$ is zero. The remaining words, composed solely of
$Z$ and identity operators, each contribute their coefficients to the final
expectation value.

An explicit example of an exact propagation is reported in
Appendix~\ref{sec:stepbystep}.

\subsection{Explicit representation of circuits}
\label{subsec:explicit}

\begin{figure*}[!t]
	\centering
	\resizebox{.9\linewidth}{!}{\includegraphics{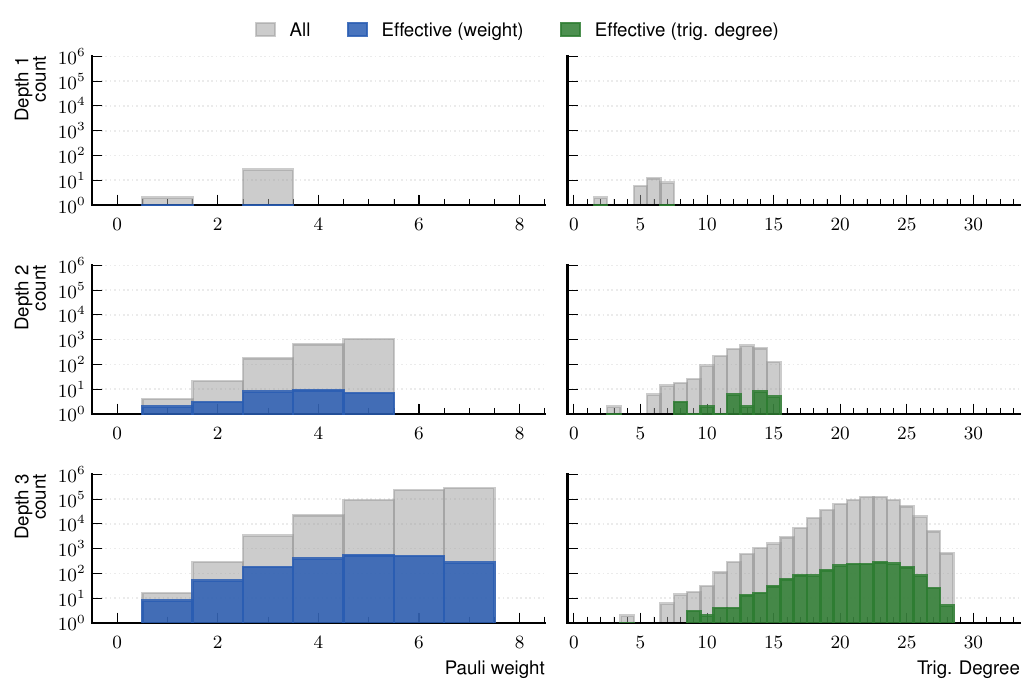}}
	\caption{Distribution of the Pauli weights (left) and trigonometric degrees (right)
		for the propagated observable $Z_{15}$ of the local entangler ansatz at
		different depths for $16$ qubits.	\label{fig:locent_distdepth} }
\end{figure*}

Before introducing the truncation schemes of Secs.~\ref{subsec:truncW} and
\ref{subsec:truncV}, we first examine how exact, untruncated propagation
scales with circuit depth. Even for a modest system size, the number of
surviving Pauli words grows so quickly that both storing them and applying
the propagation rules to each of them at every subsequent gate become
increasingly costly, motivating the cutoffs introduced later in this
section.

For this study, we consider the 16-qubit local entangler ansatz, whose circuit
representation is shown in Fig.~\ref{fig:locent_ansatz} at depths 1, 2 and 3,
corresponding to successive iterations of the blocks inside the dashed
rectangle. We focus on analyzing the propagation of the observable $Z$ on the
last qubit ($Z_{15}$) through these circuit layers.

Fig.~\ref{fig:locent_distdepth} shows the distribution of the final propagated
terms in terms of Pauli weights (left) and trigonometric degrees (right), with each row
corresponding to a different circuit depth. The colored terms (blue on the left,
green on the right) represent the effective terms that survive the trimming
process, contributing to the functional form of the observable, while the gray
terms indicate those discarded during the expectation value computation.

The plots on the left illustrate how the Pauli weights increase with circuit
depth. Specifically, at each iteration of the local entangler ansatz, any Pauli
word can increase its weight by at most two, due to the two distinct layers of
CNOT gates.

The plots on the right depict a gradual increase in the trigonometric degrees of the
terms, with the maximum possible trigonometric degree equal to the total number of
parameters. We observe an accumulation of higher-degree terms as the circuit
depth grows.

Note that all the y-axes in the plots are displayed on a logarithmic scale,
highlighting how rapidly the number of terms grows during propagation and
illustrating why the exact method becomes computationally intractable, thus
necessitating suitable cutoffs.

\begin{figure}
	\centering
	\begin{quantikz}[row sep={0.6cm,between origins}, wire types={q,q,q,n,q}]
		\lstick{$\ket{0}$}                            & \gate{RY}
		\gategroup[6,steps=4,style={dashed}, label style={label
		position=below,anchor=north,yshift=-0.2cm}]{} & \gate{RX} & \qw &
		\ctrl{1}                                      & \gate{RY} &       \\
		\lstick{$\ket{0}$}                            &
		\gate{RY}
		                                              &
		\gate{RX}
		                                              &
		\ctrl{1}
		                                              &
		\targ{1}
		                                              &
		\gate{RY}
		                                              &
		\\
		\lstick{$\ket{0}$}                            &
		\gate{RY}
		                                              &
		\gate{RX}
		                                              &
		\targ{1}
		                                              &
		\ctrl{0}
		                                              &
		\gate{RY}
		                                              &
		\\
		\lstick{\,\vdots}                             & \,\vdots  &     &
		\vdots
		                                              &
		\vdots
		                                              &           &
		\\
		\lstick{$\ket{0}$}                            &
		\gate{RY}
		                                              &
		\gate{RX}
		                                              &
		\targ{0}
		                                              &
		\ctrl{1}
		                                              &
		\gate{RY}
		                                              &
		\\
		\lstick{$\ket{0}$}                            &
		\gate{RY}
		                                              &
		\gate{RX}
		                                              &           &
		\targ{1}
		                                              &
		\gate{RY}
		                                              &
	\end{quantikz}
	\caption{Local entangler Ansatz}
	\label{fig:locent_ansatz}
\end{figure}
\subsection{Truncation on Pauli Weight}
\label{subsec:truncW}

During propagation, multi-qubit gates induce an increase in the weight of the
propagated Pauli words. The work proposed in Ref. \cite{classicallyestimating}
showed that, on average, propagated observables from locally scrambling ansätze
contain terms whose importance decreases exponentially with increasing weight.
This property allows setting a cutoff value \( w_{\mathrm{cut}} \) during
propagation, discarding any words that exceed this cutoff and preventing them
from further branching. This significantly reduces the computational burden.

\subsection{Truncation on Trigonometric Degree}
\label{subsec:truncV}
Other than the weights, additional terms during propagation can be reasonably
discarded without significantly impacting the accuracy of the observables.
During propagation, at each non-commuting parameterized gate acting on a Pauli
word, that word branches and introduces either the sine or cosine of the
corresponding parameter, increasing the trigonometric degree. This results in words having
coefficients that are products of sines and cosines of various elements of
$\vec{\theta}$.

Since all sine and cosine values have absolute values less than or equal to 1,
high-degree words tend to contribute less on average because their
coefficients are more likely to be close to zero. A more rigorous argument can
be found in Appendix~\ref{sec:highfreq}.

Hence, by setting a cutoff value on the trigonometric degree, $\delta_\mathrm{cut}$, we can
discard during propagation all those words whose trigonometric degree exceeds this
threshold.

In practice, the two truncation schemes are applied jointly, discarding any
Pauli word whose weight exceeds $w_{\mathrm{cut}}$ or whose trigonometric degree exceeds
$\delta_{\mathrm{cut}}$. Under a decay assumption on the propagated coefficients
consistent with locally scrambling ansätze, this joint truncation can be shown
to introduce an error on the expectation value that decays exponentially in both
$w_{\mathrm{cut}}$ and $\delta_{\mathrm{cut}}$, with an analogous guarantee holding
for the corresponding gradients. The full derivation and the resulting bound are
provided in Appendix~\ref{sec:jamal_truncation}.

\section{Results}
\label{sec:res}
The ability to obtain a fast, approximate functional description of observables
for a given circuit can be readily exploited in various \gls{qml} algorithms
that involve the minimization of a cost function dependent on one or more
observables. Among these, the \gls{vqe}~\cite{vqe} stands out as one of the most
well-known examples, and an application of this approach is presented in the
following subsection.
\subsection{Variational Quantum Eigensolver}
\label{subsec:vqe}
The \gls{vqe} is a quantum algorithm designed to approximate the ground state of
a given Hamiltonian $H$. For a parameterized quantum circuit $U(\vec{\theta})$,
the variational state is defined as $|\psi(\vec{\theta})\rangle =
	U(\vec{\theta})|0\rangle$. The objective is to find the set of parameters
$\vec{\theta}$ that minimizes the expectation value of the Hamiltonian, which
corresponds to the energy
\begin{equation}
	\mathcal{L}(\vec{\theta}) = \langle \psi(\vec{\theta}) | H | \psi(\vec{\theta}) \rangle,
\end{equation}
such that the evolved state $|\psi(\vec{\theta})\rangle$ approximates the true
ground state of $H$ as closely as possible.

In this case, the observable to be propagated is the model's Hamiltonian itself,
whose expectation value defines the loss function. The resulting variational
state $|\psi(\vec{\theta})\rangle$ obtained after optimization can then be
employed as an input to other \gls{qml} models~\cite{Monaco_2023, ad}, serving
as true quantum input data.

To illustrate an application of the symbolic propagation, the \gls{annni} model
is used as an example, whose Hamiltonian is defined as
\begin{equation}
	\label{eq:annni_ham}
	H(\kappa, h) = -J \sum_{i=1}^{N} \left( Z_i Z_{i+1} - \kappa Z_i Z_{i+2} + h X_i \right),
\end{equation}
where $J$ denotes the nearest-neighbor coupling constant (set to $J = 1$ without
loss of generality), $\kappa$ controls the strength of the next-nearest-neighbor
interaction, and $h$ represents the transverse magnetic field strength.

Depending on the values of $\kappa$ and $h$, the system exhibits different
phases arising from the competition between the transverse magnetic field and
the two types of spin-spin interactions. Since Eq.~\ref{eq:annni_ham} is linear
in $\kappa$ and $h$, regardless of their specific values, the Hamiltonian can be
decomposed into three distinct observables, $O_1 = \sum_{i=1}^{N} X_i X_{i+1}$,
$O_2 = \sum_{i=1}^{N} X_i X_{i+2}$, and $O_3 = \sum_{i=1}^{N} Z_i$, which only
need to be propagated once. Consequently, a single propagation of these three
observables is sufficient to obtain the expectation value of the Hamiltonian
$H(\kappa, h)$ for any value of $(\kappa, h)$ and any choice of parameters
$\vec{\theta}$, for the given propagated circuit.

Once the explicit functional form of the Hamiltonian in terms of the parameters
is obtained, the gradient of the loss function with respect to these parameters
can be readily computed either analytically or via the \textit{parameter-shift}
rule~\cite{paramshift}, as both approaches yield the same gradient function
(Appendix \ref{sec:paramshift}).

\begin{figure}[ht]
	\centering
	\includegraphics{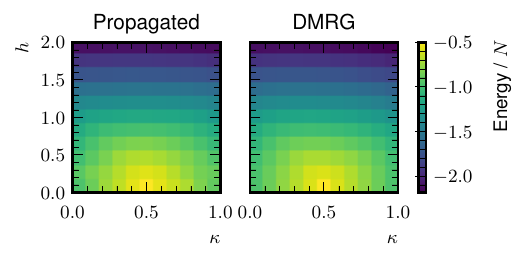}\vspace*{-.5cm}
	\caption{Energies of
		the 32-spin ANNNI model at various values of $\kappa$ and $h$, obtained
		using \gls{dmrg} (right) and by minimizing the function of the propagated
		Hamiltonian via gradient descent (left).}
	\label{fig:vqephasediagE}
	\vspace*{.3cm}
	\includegraphics{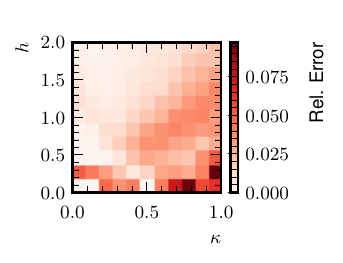}\vspace*{-.5cm}
	\caption{Relative Errors of the energies between the VQE energies and DMRG
		energies from Fig. \ref{fig:vqephasediagE}}
	\label{fig:vqephasediagRelerr}
\end{figure}

A simulation was carried out on a system of $32$ qubits with $3$ iterations of
the local entangler ansatz introduced in Fig.~\ref{fig:locent_ansatz}. The
optimization was performed using the Adam optimizer~\cite{adam} on the
propagated observables, minimizing the expectation value of the three
observables $O_1$, $O_2$, and $O_3$. While larger system sizes and deeper
circuits can, in principle, be treated with this method, we chose this
configuration so that we could still run the optimization without any
truncation, allowing us to obtain the exact expectation values rather than their
approximated counterparts. The results of these simulations are presented in
Fig.~\ref{fig:vqephasediagE}.

Figure~\ref{fig:vqephasediagE} presents a comparison between the ground-state
energies of the \gls{annni} model obtained through
\gls{dmrg}~\cite{schollwock2005density,orus2019tensor} (left) and those computed
via the \gls{vqe} using the propagated Hamiltonian (right), while
Fig.~\ref{fig:vqephasediagRelerr} reports the corresponding relative errors
between the two results. Overall, the training achieves high accuracy with the
most significant deviations being observed in regions characterized by low
values of $h$ and high values of $\kappa$, corresponding to the so-called
\textit{antiphase} of the model, where the spin configuration exhibits a less
regular and more non-local ordering pattern.

\begin{figure}[ht]
	\centering
	\subfloat[VQE training with fixed $\delta_{\mathrm{cut}} = 20$ and varying
		$w_{\mathrm{cut}} \in \{2, 4, 8\}$. \label{subfig:lines_w}]{%
		\includegraphics{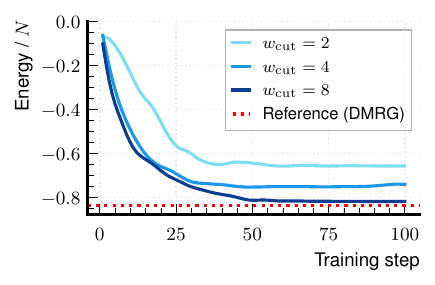}
	}\hfill \subfloat[VQE training with fixed $w_{\mathrm{cut}} = 8$ and varying
		$\delta_{\mathrm{cut}} \in \{10, 15, 20\}$. \label{subfig:lines_nu}]{%
		\includegraphics{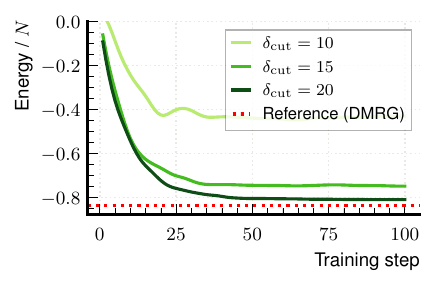}
	}
	\caption{Training curves of the VQE applied to the ANNNI Hamiltonian at point $(\kappa, h) = (0.2, 0.4)$.}
	\label{fig:vqeline}
\end{figure}

\paragraph{Effects of Truncation on Training}
To investigate the effect of the truncation parameters on the training,
Fig.~\ref{fig:vqeline} shows the training curves for a representative point of
the phase diagram, $(\kappa, h) = (0.2, 0.4)$, obtained for various cutoff
values of $w_\mathrm{cut}$ and $\delta_\mathrm{cut}$. Increasing $w_\mathrm{cut}$
leads to an immediate improvement in accuracy, with $w_\mathrm{cut}=8$ yielding
a result close to the reference value. A similar trend is observed for
$\delta_\mathrm{cut}$, where a relatively small value of $\delta_\mathrm{cut}=20$
already provides near-perfect convergence, highlighting the efficiency and
simulability of the proposed approximation. Table~\ref{tab:num_obs} reports the
number of Pauli words retained under each of these cutoffs applied individually.
Notably, $\delta_{\mathrm{cut}}=20$ already discards over $90\%$ of the terms while
achieving results close to the reference.

\paragraph{Scalability}
Table~\ref{tab:num_obs_L} reports the total number of retained Pauli words as a
function of the system size $N$, with no truncation applied, together with the
corresponding number of words per Hamiltonian term. While the total number of
Pauli words necessarily grows with $N$, since a larger system corresponds to a
proportionally larger Hamiltonian, the number of words per Hamiltonian term
saturates rapidly, growing from $1225$ at $N=4$ to $14326$ at $N=64$, with the
largest relative increase occurring between $N=4$ and $N=8$ and comparatively
little change for $N \geq 16$. This behavior follows directly from the locally
scrambling nature of the ansatz: as discussed in Sec.~\ref{subsec:explicit},
each layer grows a propagated Pauli word's support by a bounded amount, so after
$d$ layers it is confined to a \emph{light cone} whose radius is set by the
depth $d$, not by $N$; this bounded per-layer growth is visible in
Fig.~\ref{fig:locent_distdepth}, where the Pauli-weight distribution of $Z_{15}$
advances by at most two units per layer. Under this ansatz the light cone
eventually stops growing once it saturates the locality bound, so the
propagation never reaches beyond qubit $8$, however large $N$ is. Adding qubits
beyond this light cone therefore does not raise the propagation cost of any
individual term. The real cost driver is instead the depth $d$: increasing $d$
enlarges the light cone and, with it, the number of Pauli words that must be
tracked per observable, which is what eventually saturates available memory. The
polynomial scaling of the overall cost with $N$ comes from a different source:
the number of initial observables, one per Hamiltonian term, grows polynomially
with $N$, while the propagation cost of each individual observable saturates to
a constant once $N$ exceeds its light-cone radius. This makes the method best
suited to relatively shallow ansätze rather than deep ones.
\begin{table}[h]
	\centering
	\begin{minipage}{0.45\linewidth}
		\centering
		\begin{tabular}{@{}rr@{}}
			\hline
			$\delta_{\mathrm{cut}}$ & \rule{.5cm}{0pt} Num. Pauli words \\
			\hline
			10                      & 420                               \\
			15                      & 9237                              \\
			20                      & 89731                             \\
			-                       & 1257590                           \\
			\hline
		\end{tabular}
	\end{minipage}\hspace*{.5cm}
	\begin{minipage}{0.45\linewidth}
		\centering
		\begin{tabular}{@{}rr@{}}
			\hline
			$w_{\mathrm{cut}}$ & \rule{.5cm}{0pt} Num. Pauli words \\
			\hline
			2                  & 311                               \\
			4                  & 44008                             \\
			8                  & 1243622                           \\
			-                  & 1257590                           \\
			\hline
		\end{tabular}
	\end{minipage}
	\caption{Number of Pauli words retained as a function of the $\delta$ and $w$ truncation cutoffs.}
	\label{tab:num_obs}
\end{table}

\begin{table}[h]
	\centering
	\begin{tabular}{@{}rrr@{}}
		\hline
		$N$ & \rule{.5cm}{0pt}\makecell{Num. Pauli Words} &
		\rule{.5cm}{0pt}\makecell{Num. Pauli words /                  \\
		Hamiltonian}                                                  \\
		\hline
		4   & 11026                                       & 1225.111  \\
		8   & 170078                                      & 8098.952  \\
		16  & 532582                                      & 11835.156 \\
		32  & 1257590                                     & 13522.473 \\
		64  & 2707606                                     & 14325.958 \\
		\hline
	\end{tabular}
	\caption{Number of Pauli words retained as a function of the system size $N$, with no $\delta$ or $w$ truncation.}
	\label{tab:num_obs_L}
\end{table}

\section{Discussions and conclusions}
\label{sec:conc}
The \gls{vqe} represents only one class of problems that can benefit from
pre-training through the proposed propagation-based approach. Nevertheless, it
remains a particularly promising example, as locally scrambling ansätze align
naturally with the description of spin models, which, under most formulations,
are governed by local interactions. Other \gls{qml} tasks can similarly leverage
this method, such as compression via anomaly-detection ansätze~\cite{ad} and
bitstring generation through the formulation of the \gls{mmd} loss as randomly
sampled observables in the $Z$ basis~\cite{trainabilitymmd}. However, in the
latter case, the number of observables to be propagated can become prohibitively
large, and the approximate training of the observable estimator may introduce
excessive error, limiting the method's applicability.

This approach enables an effective form of off-chip pre-training, thereby
alleviating the computational burden associated with costly on-chip quantum
optimization. For such pre-training to be applicable, two main conditions must
be satisfied: (i) the problem must be expressed through a loss function defined
in terms of observables, and (ii) the chosen ansatz must exhibit favorable
propagation properties, most notably, locally scrambling that ensures efficient
and accurate observable evaluations.

Future developments may focus on extending this framework along three main
directions. First, improvements to the propagation routine could allow deeper
circuits to be propagated more efficiently. Second, this method could be used to
study ansatz construction. Third, this approach could be broadened to other
classes of \gls{qml} problems beyond the \gls{vqe} setting.

\section{Acknowledgments}
This research was supported in part through the Maxwell computational resources
operated at Deutsches Elektronen-Synchrotron DESY (Hamburg, Germany), a member
of the Helmholtz Association (HGF).\\
ENGAGE has received funding from the European Union's Horizon 2020 Research and
Innovation Programme under the Marie Skłodowska-Curie Grant Agreement No.
101034267.\\
We thank Ran Xue for useful conversations and suggestions on this work and
its code.

\section{Code availability}
All code developed and used for the analyses presented in this work is publicly
available on GitHub at
\url{https://github.com/SaverioMonaco/Pauli-Propagator}~\cite{githubrepo}.

\nocite{pennylane,kay2018tutorial}

\bibliography{references}

\appendix
\section{Step-by-step propagation}
\label{sec:stepbystep}
To illustrate how the Pauli propagation of observables works, we consider as an
example the circuit shown in Fig.~\ref{fig:4locent_ansatz}, focusing on the
propagation of the observable $Z_0$ through the layers of the ansatz.

\begin{figure}[ht]
	\centering
	\newcommand{\gatewidth}{1.2cm}
	\begin{quantikz}[row sep={0.6cm,between origins}, wire types={q,q,q,q}]
		\lstick{$\ket{0}$}                       & \gate[][\gatewidth]{RY(\theta_0)} &
		\ctrl{1}                                 &
		\gate[][\gatewidth]{RX(\theta_4)}        & \qw                               &
		\gate[][\gatewidth]{RY(\theta_8)}        &                                     \\
		\lstick{$\ket{0}$}                       & \gate[][\gatewidth]{RY(\theta_1)} &
		\targ{1}                                 &
		\gate[][\gatewidth]{RX(\theta_5)}        & \ctrl{1}                          &
		\gate[][\gatewidth]{RY(\theta_9)}        &                                     \\
		\lstick{$\ket{0}$}                       & \gate[][\gatewidth]{RY(\theta_2)} &
		\ctrl{1}                                 &
		\gate[][\gatewidth]{RX(\theta_6)}        & \targ{1}                          &
		\gate[][\gatewidth]{\!RY(\theta_{10})\!} &                                     \\
		\lstick{$\ket{0}$}                       & \gate[][\gatewidth]{RY(\theta_3)} &
		\targ{1}                                 &
		\gate[][\gatewidth]{RX(\theta_7)}        & \qw                               &
		\gate[][\gatewidth]{\!RY(\theta_{11})\!} &
	\end{quantikz}
	\caption{Four-qubit local entangler ansatz with a single iteration (depth = 1).}
	\label{fig:4locent_ansatz}
\end{figure}

\begin{table}[h]
	\centering
	\setlength{\tabcolsep}{0pt}
	\renewcommand{\arraystretch}{1.2}

	\rule{.8\linewidth}{.3pt}
	\begin{tabular}{@{}l@{\hspace{1cm}}l@{}}
		\begin{quantikz}[row sep=0.2cm, column sep=0.1cm]
			\qw & \gate{RX(\theta)} & \qw
		\end{quantikz}
		 &
		\(
		\begin{array}{rcl}
			I & \longrightarrow & I                           \\
			X & \longrightarrow & X                           \\
			Y & \longrightarrow & \cos\theta Y - \sin\theta Z \\
			Z & \longrightarrow & \cos\theta Z + \sin\theta Y
		\end{array}
		\)
	\end{tabular}

	\rule{.8\linewidth}{.3pt}

	\begin{tabular}{@{}l@{\hspace{1cm}}l@{}}
		\begin{quantikz}[row sep=0.2cm, column sep=0.1cm]
			\qw & \gate{RY(\theta)} & \qw
		\end{quantikz}
		 &
		\(
		\begin{array}{rcl}
			I & \longrightarrow & I                           \\
			X & \longrightarrow & \cos\theta X + \sin\theta Z \\
			Y & \longrightarrow & Y                           \\
			Z & \longrightarrow & \cos\theta Z - \sin\theta X
		\end{array}
		\)
	\end{tabular}

	\rule{.8\linewidth}{.3pt}

	\begin{tabular}{@{}l@{\hspace{1cm}}l@{}}
		\begin{quantikz}[row sep=0.2cm, column sep=0.1cm]
			\qw & \ctrl{1} & \qw \\
			\qw & \targ{}  & \qw
		\end{quantikz}
		 &
		\(
		\begin{array}{rcl}
			I \otimes I & \longrightarrow & I \otimes I \\
			X \otimes I & \longrightarrow & X \otimes X \\
			Y \otimes I & \longrightarrow & Y \otimes X \\
			Z \otimes I & \longrightarrow & Z \otimes I \\
			I \otimes X & \longrightarrow & I \otimes X \\
			I \otimes Y & \longrightarrow & Z \otimes Y \\
			I \otimes Z & \longrightarrow & Z \otimes Z \\
			X \otimes X & \longrightarrow & X \otimes I \\
			Y \otimes X & \longrightarrow & Y \otimes I \\
			Z \otimes Y & \longrightarrow & I \otimes Y \\
			Z \otimes Z & \longrightarrow & I \otimes Z
		\end{array}
		\)
	\end{tabular}
	\rule{.8\linewidth}{.3pt}

	\caption{Propagation rules for the $R_X(\theta)$, $R_Y(\theta)$, and CNOT gates.}
	\label{tab:ops}
\end{table}
\begin{itemize}
	\item \textbf{Layer 5:} \mbox{($RY(\theta_8)_0\, RY(\theta_9)_1\,
			      RY(\theta_{10})_2\, RY(\theta_{11})_3$)}

	      At this stage, the observable $Z_0$ is only affected by the gate
	      $RY(\theta_8)_0$. According to the propagation rules in
	      Table~\ref{tab:ops}, this evolves as:
	      \begin{equation}
		      Z_0 \to \cos\theta_8 Z_0 - \sin\theta_8X_0
	      \end{equation}

	\item \textbf{Layer 4:} \mbox{($\text{CNOT}(1,2)$)}

	      Both terms of the observable commute with $\text{CNOT}(1,2)$, hence no
	      transformation occurs:
	      \begin{equation}
		      \cos\theta_8 Z_0 - \sin\theta_8X_0 \to \cos\theta_8 Z_0 - \sin\theta_8X_0
	      \end{equation}
	\item \textbf{Layer 3:} \mbox{($RX(\theta_4)_0\, RX(\theta_5)_1\,
			      RX(\theta_6)_2\, RX(\theta_7)_3$)}

	      Here only the gate $RX(\theta_4)_0$ affects the propagated observable.
	      Using the propagation rules for $RX(\theta)$, we obtain:
	      \begin{align}
		      \cos\theta_8 Z_0 - \sin\theta_8X_0 \to & + \cos\theta_8\cos\theta_4 Z_0\notag  \\
		                                             & + \cos\theta_8\sin\theta_4 Y_0 \notag \\
		                                             & - \sin\theta_8X_0
	      \end{align}
	\item \textbf{Layer 2:} \mbox{($\text{CNOT}(0,1)\,\text{CNOT}(2,3)$)}

	      The $\text{CNOT}(0,1)$ gate acts on all observables involving the
	      first two qubits. Applying the propagation rules for the CNOT yields:
	      \begin{align}
		      + & \cos\theta_8\cos\theta_4 Z_0 + \cos\theta_8\sin\theta_4 Y_0 - \sin\theta_8X_0                   \notag \\
		        & \downarrow                                                              \notag                         \\
		        & + \cos\theta_8\cos\theta_4 Z_0 + \cos\theta_8\sin\theta_4 Y_0X_1 \notag                                \\
		        & - \sin\theta_8X_0X_1
	      \end{align}
	\item \textbf{Layer 1:} \mbox{($RY(\theta_0)_0\, RY(\theta_1)_1\,
			      RY(\theta_{2})_2\, RY(\theta_{3})_3$)}

	      In this final layer, both $RY(\theta_0)_0$ and $RY(\theta_1)_1$
	      contribute to the evolution.

	      \begin{itemize}

		      \item \textbf{Action of $RY(\theta_1)_1$:}
		            \begin{align}
			            + & \cos\theta_8\cos\theta_4 Z_0 + \cos\theta_8\sin\theta_4 Y_0X_1 \notag                              \\
			            - & \sin\theta_8X_0X_1 \notag                                                                          \\
			              & \downarrow                                                              \notag                     \\
			              & + \cos\theta_8\cos\theta_4 Z_0 \notag                                                              \\
			              & + \cos\theta_8\sin\theta_4\cos\theta_1 Y_0X_1 - \cos\theta_8\sin\theta_4\sin\theta_1 Y_0Z_1 \notag \\
			              & - \sin\theta_8\cos\theta_1X_0X_1 + \sin\theta_8\sin\theta_1X_0Z_1
		            \end{align}
		      \item \textbf{Action of $RY(\theta_0)_0$:}
		            \begin{align}
			            + & \cos\theta_8\cos\theta_4 Z_0 \notag                                                                \\
			            + & \cos\theta_8\sin\theta_4\cos\theta_1 Y_0X_1 - \cos\theta_8\sin\theta_4\sin\theta_1 Y_0Z_1 \notag   \\
			            - & \sin\theta_8\cos\theta_1X_0X_1 + \sin\theta_8\sin\theta_1X_0Z_1 \notag                             \\
			              & \downarrow                                                              \notag                     \\
			              & + \cos\theta_8\cos\theta_4\cos\theta_0 Z_0 - \cos\theta_8\cos\theta_4\sin\theta_0 X_0 \notag       \\
			              & + \cos\theta_8\sin\theta_4\cos\theta_1 Y_0X_1 - \cos\theta_8\sin\theta_4\sin\theta_1 Y_0Z_1 \notag \\
			              & - \sin\theta_8\cos\theta_1\cos\theta_0X_0X_1 - \sin\theta_8\cos\theta_1\sin\theta_0Z_0X_1 \notag   \\
			              & + \sin\theta_8\sin\theta_1\cos\theta_0X_0Z_1 - \sin\theta_8\sin\theta_1\sin\theta_0Z_0Z_1 \notag   \\
		            \end{align}
	      \end{itemize}
	\item \textbf{Expectation value computation (trimming):}

	      When computing the expectation value, each propagated observable is
	      sandwiched between $\langle 0|$ and $|0\rangle$. Only Pauli strings
	      composed solely of $Z$ or $I$ operators survive this projection,
	      leading to:
	      \begin{align}
		      \langle Z_0\rangle(\vec\theta) = - \sin\theta_8\sin\theta_1\sin\theta_0 + \cos\theta_8\cos\theta_4\cos\theta_0
	      \end{align}
\end{itemize}
\section{Contribution of High-Degree Terms}
\label{sec:highfreq}

When the training process reaches an optimal set of parameters~$\{\theta_i\}$,
the individual values of~$\theta_i$ are typically correlated as a result of the
optimization. Nevertheless, to obtain an a priori estimate of the contribution
of high-degree terms, independent of any specific problem instance, it is
convenient to first treat the parameters~$\theta_i$ as random and independent;
the validity of this simplifying assumption once training has converged is
examined later in this section.

For simplicity, we assume that the parameters are uniformly distributed,
$\theta_i \sim U[0,2\pi]$. Let $f_i(\theta_i)$ denote either $\sin\theta_i$ or
$\cos\theta_i$. Under this assumption,
\[
	\mathbb{E}[f_i(\theta_i)] = 0,
	\qquad
	\mathbb{E}[f_i^2(\theta_i)] = \frac{1}{2}.
\]
For a product of $\delta$ such factors,
\[
	\mathbb{E}\!\left(\prod_{i=1}^{\delta} f_i^2(\theta_i)\right)
	= \prod_{i=1}^{\delta} \mathbb{E}[f_i^2(\theta_i)]
	= \frac{1}{2^{\delta}}.
\]

This result shows that each additional sine or cosine factor contributes a fixed
multiplicative reduction to the overall variance. More generally, even for
broader classes of distributions with zero mean and bounded support, the
variance of the product decays exponentially with the number of factors~$\delta$.
Hence, as the string of trigonometric functions grows longer, its expected value
becomes increasingly concentrated around zero, justifying the truncation of
high-degree terms.

\begin{figure}[ht]
	\centering
	\includegraphics{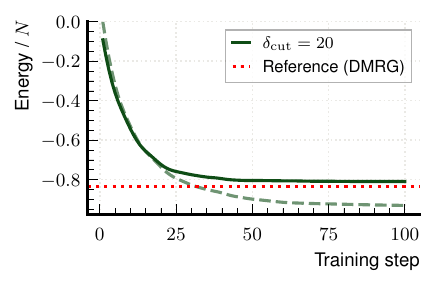}\vspace*{-.5cm} \caption{Training run
		at $\delta_{\mathrm{cut}}=20$ (cf. Fig.~\ref{subfig:lines_nu}): exact,
		untruncated energy evaluated at the trained parameters at each step (solid)
		against the truncated energy actually minimized by the optimizer (dashed),
		compared to the DMRG ground-state reference (dotted).}
	\label{fig:truncnugoescrazy}
\end{figure}

As anticipated above, however, this assumption need not hold once training has
converged: gradient-based optimization actively correlates the parameters as it
minimizes the loss, and can in principle steer the discarded high-degree
terms away from their random-parameter average.
Figure~\ref{fig:truncnugoescrazy} illustrates this effect for one of the
training runs of Fig.~\ref{subfig:lines_nu}, at $\delta_{\mathrm{cut}}=20$.

During training, the optimizer only has access to the truncated (dashed) energy,
and its deviation from the exact (solid) energy grows as training progresses,
confirming that the parameters are no longer well described as random once
optimization has taken effect. Nevertheless, because the truncated loss is still
dominated by the same low-degree terms that dominate the exact loss,
minimizing it continues to steer the parameters towards a good solution: the
exact energy converges close to the true ground-state energy, consistent with
the variational principle, which guarantees it can never fall below it. The
truncated energy, in contrast, is not itself a variational quantity and is free
to drop below this bound, as it does here, since the terms discarded at
$\delta_{\mathrm{cut}}=20$ would, if retained, have pushed its value back above the
true ground-state energy.

\section{Equivalence between gradients and the parameter-shift rule}
\label{sec:paramshift}

The parameter-shift rule provides an exact method to compute derivatives of
expectation values with respect to circuit parameters~\cite{paramshift}. For an
observable $O(\vec{\theta})$, the derivative with respect to the parameter
$\theta_i$ is given by
\begin{equation}
	\frac{\partial O(\vec{\theta})}{\partial \theta_i}
	=
	\frac12 \left[
		\langle O\rangle_{\theta_i + \frac\pi2}
		-
		\langle O\rangle_{\theta_i - \frac\pi2}
		\right] .
	\label{eq:parameter-shift-rule}
\end{equation}

Consider now an observable truncated at Pauli weight $w_{\mathrm{cut}}$ and
trigonometric degree $\delta_{\mathrm{cut}}$, denoted by $O^{\mathrm{cut}}(\vec{\theta})$. Its
expectation value can be expressed, with respect to a single parameter
$\theta_i$, as
\begin{equation}
	\langle O^{\mathrm{cut}}(\vec{\theta}) \rangle
	=
	\sin(\theta_i)\, A_i
	+
	\cos(\theta_i)\, B_i
	+
	C_i ,
\end{equation}
where $A_i$, $B_i$, and $C_i$ are independent of $\theta_i$. This decomposition
reflects the fact that, after truncation, each retained term in the observable
either depends on $\theta_i$ through a sine or cosine factor, or is independent
of $\theta_i$ altogether.

Taking the derivative of the above expression yields
\begin{equation}
	\frac{\partial \langle O^{\mathrm{cut}}(\vec{\theta}) \rangle}{\partial \theta_i}
	=
	\cos(\theta_i)\, A_i
	-
	\sin(\theta_i)\, B_i .
	\label{eq:parameter-shift-rule-grad}
\end{equation}

We now show that this result is exactly reproduced by the parameter-shift rule
in Eq.~\ref{eq:parameter-shift-rule}. Evaluating the shifted expectation values,
we obtain
\begin{align}
	\frac12 \Big[
	\langle O & \rangle_{\theta_i + \frac\pi2}
		-
		\langle O \rangle_{\theta_i - \frac\pi2}
		\Big]
	=   \notag                                 \\
	          & = \frac12 \Big[
	+ \sin\!\left(\theta_i + \frac\pi2\right) A_i
	+
	\cos\!\left(\theta_i + \frac\pi2\right) B_i
	+
	C_i
	\notag                                     \\
	          & \hphantom{= \frac12 \Big[}
		-
		\sin\!\left(\theta_i - \frac\pi2\right) A_i
		-
		\cos\!\left(\theta_i - \frac\pi2\right) B_i
		-
		C_i
		\Big]
	\notag                                     \\
	          & =
	\cos(\theta_i)\, A_i
	-
	\sin(\theta_i)\, B_i .
\end{align}

This expression coincides exactly with the analytical gradient in
Eq.~\ref{eq:parameter-shift-rule-grad}, thereby establishing the equivalence
between the direct differentiation of the truncated observable and the
parameter-shift rule.

\section{Error bound for joint Pauli-weight and trigonometric-degree truncation}
\label{sec:jamal_truncation}

In Secs. \ref{subsec:truncW} and \ref{subsec:truncV} we introduced the
truncation on Pauli weight $w_{\mathrm{cut}}$ and on trigonometric degree
$\delta_{\mathrm{cut}}$. We now derive a simple, explicit upper bound on the error
induced by this joint truncation.

Recall from Sec. \ref{subsec:exact} that, after symbolic propagation and
trimming on $\lvert 0\rangle^{\otimes n}$, any propagated observable
$O(\vec{\theta})$ can be written as a finite sum of Pauli words with
coefficients that are products of sines and cosines of the components of
$\vec{\theta}$ as

\begin{equation}
	O(\vec{\theta})=\sum_{j} c_j(\vec{\theta})\, P_j,
	\label{eq:O-symbolic-decomposition}
\end{equation}
with
\begin{equation}
	c_j(\vec{\theta})
	= \prod_{\ell=1}^{\delta_j} f_{j,\ell}(\theta_{i_{j,\ell}}),
	\label{eq:O-symbolic-decomposition-coefficents}
\end{equation}

where each $P_j$ is a Pauli word, $\delta_j$ is the trigonometric degree of that term (as
defined in Sec. \ref{subsec:explicit}, i.e.\ the number of sine/cosine factors),
and each $f_{j,\ell}$ is either $\sin$ or $\cos$ of one of the parameters. We
write $w_j := \mathrm{weight}(P_j)$ for the Pauli weight. Let $L(\vec{\theta}) =
	\langle \psi(\vec{\theta})|O|\psi(\vec{\theta})\rangle$ be the loss function,
since $\lvert\langle 0^{\otimes n} | P_j | 0^{\otimes n}\rangle\rvert \le 1,
	\forall P_j$, we can bound the total contribution of each term by

\begin{equation}
	\big| c_j(\vec{\theta})\,\langle 0^{\otimes n} | P_j | 0^{\otimes n}\rangle \big|\le \big|c_j(\vec{\theta})\big|.
	\label{eq:coeff-overlap-bound}
\end{equation}

By-design, each factor satisfies $|f_{j,\ell}(\cdot)| \le 1$, so that
$|c_j(\vec{\theta})| \le 1$ for all $\vec{\theta}$. This trivial uniform bound,
however, does not control how the truncation error scales with
$w_{\mathrm{cut}}$ and $\delta_{\mathrm{cut}}$. To obtain a meaningful scaling
bound, we instead promote the locality and scrambling intuition to an explicit
decay assumption on the coefficients.

\begin{assumption}[Weight-degree decay]\label{ass:weight-frequency-decay}
	There exist constants $C_0 > 0$ and $0 < \alpha,\beta < 1$ such that, for
	any term $j$ with Pauli weight $w_j$ and trigonometric degree $\delta_j$, its coefficient
	satisfies
	\begin{equation}
		\sup_{\vec{\theta}} |c_j(\vec{\theta})| \le C_0\, \alpha^{w_j}\, \beta^{\delta_j}.
		\label{eq:decay-assumption}
	\end{equation}
\end{assumption}

This captures the key idea that, for locally scrambling circuits at fixed depth
and bounded-degree connectivity, large-weight and high-degree terms appear
only through many small contributions and are exponentially suppressed on
average.

Let $O_{w_{\mathrm{cut}},\delta_{\mathrm{cut}}}(\vec{\theta})$ denote the truncated
observable obtained by discarding all terms with $w_j>w_{\mathrm{cut}}$ or
$\delta_j > \delta_{\mathrm{cut}}$ during propagation, and let
$L_{w_{\mathrm{cut}},\delta_{\mathrm{cut}}}(\vec{\theta})$ be the corresponding
truncated loss. In the worst-case truncation error

\begin{equation}
	\Delta_{w_{\mathrm{cut}},\delta_{\mathrm{cut}}}(\vec{\theta})
	:=
	L(\vec{\theta}) - L_{w_{\mathrm{cut}},\delta_{\mathrm{cut}}}(\vec{\theta}).
	\label{eq:Delta-def}
\end{equation}

By construction, this difference is a sum over precisely those terms with $w_j >
	w_{\mathrm{cut}}$ or $\delta_j > \delta_{\mathrm{cut}}$.

We introduce the sets
\begin{equation}
	\mathcal{J}_w := \{ j : w_j > w_{\mathrm{cut}} \}, \quad \mathcal{J}_\delta := \{ j : \delta_j > \delta_{\mathrm{cut}} \}
	\label{eq:jdef}
\end{equation}
\begin{equation}
	\mathcal{J}_{\mathrm{cut}} := \mathcal{J}_w \cup \mathcal{J}_\delta
	\label{eq:jdefjoint}
\end{equation}

Using Eq.~\ref{eq:coeff-overlap-bound} and
Assumption~\ref{ass:weight-frequency-decay}, we obtain

\begin{align}
	\sup_{\vec{\theta}}
	\big|\Delta_{w_{\mathrm{cut}},\delta_{\mathrm{cut}}}(\vec{\theta})\big|
	\le &
	\sum_{j\in\mathcal{J}_{\mathrm{cut}}}
	\sup_{\vec{\theta}} |c_j(\vec{\theta})| \notag \\
	\le &
	\sum_{j\in\mathcal{J}_{\mathrm{cut}}}
	C_0\, \alpha^{w_j}\, \beta^{\delta_j}.
	\label{eq:Delta-sup-sum-j}
\end{align}

To make this explicit, we bound how many terms can have a given $(w,\delta)$. For
each weight $w$, the number of Pauli words on $n$ qubits with $\mathrm{wt}(P)=w$
is

\begin{equation}
	N_w = \binom{n}{w} 3^w \le (3n)^w.
	\label{eq:Nw-upper}
\end{equation}

For each trigonometric degree $\delta$, the number of distinct products of $\delta$ sine or
cosine factors over $P$ parameters is at most $(2P)^\delta$ (there are $P$ choices
of parameter and two possible trigonometric functions for each factor), so we
write

\begin{equation}
	M_\delta \;\le\; (2P)^\delta.
	\label{eq:Mnu-upper}
\end{equation}

Thus, the total contribution from all terms with weight $w$ and trigonometric degree $\delta$
is bounded by

\begin{equation}
	C_0\, \alpha^w \beta^\delta\, N_w M_\delta
	\;\le\;
	C_0\, (3n\alpha)^w\, (2P\beta)^\delta.
	\label{eq:contribution-w-nu}
\end{equation}

Summing over all discarded $(w,\delta)$ therefore yields
\begin{align}
	\sup_{\vec{\theta}}
	\big|\Delta_{w_{\mathrm{cut}},\delta_{\mathrm{cut}}}(\vec{\theta})\big|
	 & \le
	\sum_{\substack{w,\delta: \\
	w > w_{\mathrm{cut}}\     \\[.1cm] \text{or}\ \\[.1cm] \delta > \delta_{\mathrm{cut}}}}
	C_0\, (3n\alpha)^w\, (2P\beta)^\delta.
	\label{eq:Delta-sup-double-sum}
\end{align}\vspace*{-.0cm}

Let $A := 3n\,\alpha$, and $B := 2P\,\beta$. Assuming that $A<1$ and $B<1$,
which simply encodes the fact that the effective (weight, degree) decay
dominates the combinatorial growth, we can then split the sum into a weight tail
and a degree tail as follows

\begin{align}
	\sup_{\vec{\theta}}
	\big|\Delta_{w_{\mathrm{cut}},\delta_{\mathrm{cut}}}(\vec{\theta})\big|
	 & \le
	C_0\!\!\!
	\sum_{\substack{w,\delta:                             \\ w > w_{\mathrm{cut}}}} A^w B^\delta
	+
	C_0\!\!\!
	\sum_{\substack{w,\delta:                             \\ \delta > \delta_{\mathrm{cut}}}} A^w B^\delta
	\nonumber                                             \\[.1cm]
	 & =: \varepsilon_{\mathrm{weight}}(w_{\mathrm{cut}})
	+  \varepsilon_{\mathrm{deg}}(\delta_{\mathrm{cut}}).
	\label{eq:Delta-sup-eps}
\end{align}

Both $\varepsilon_{\mathrm{weight}}$ and $\varepsilon_{\mathrm{deg}}$ are
geometric series. For the weight tail we have

\begin{align}
	\varepsilon_{\mathrm{weight}}(w_{\mathrm{cut}})
	 & = C_0\!\!\! \sum_{w=w_{\mathrm{cut}}+1}^{\infty}
	\sum_{\delta=0}^{\infty} A^w B^\delta
	= C_0\!\!\!\!\! \sum_{w=w_{\mathrm{cut}}+1}^{\infty}\!\!\!\! A^w
	\sum_{\delta=0}^{\infty} B^\delta
	\nonumber                                                    \\
	 & = C_0 \left(\frac{1}{1-B}\right)
	\left(\frac{A^{w_{\mathrm{cut}}+1}}{1-A}\right)       \notag \\
	 & = \frac{C_0}{(1-A)(1-B)}\, A^{w_{\mathrm{cut}}+1}.
	\label{eq:eps-weight-final}
\end{align}

Similarly, for the degree tail,

\begin{align}
	\varepsilon_{\mathrm{deg}}(\delta_{\mathrm{cut}})
	 & = C_0\!\!\!\! \sum_{\delta=\delta_{\mathrm{cut}}+1}^{\infty}
	\sum_{w=0}^{\infty} A^w B^\delta
	= C_0\!\!\! \sum_{\delta=\delta_{\mathrm{cut}}+1}^{\infty} \!\!\!\! B^\delta
	\sum_{w=0}^{\infty} A^w
	\nonumber                                                       \\
	 & = C_0 \left(\frac{1}{1-A}\right)
	\left(\frac{B^{\delta_{\mathrm{cut}}+1}}{1-B}\right) \nonumber  \\
	 & = \frac{C_0}{(1-A)(1-B)}\, B^{\delta_{\mathrm{cut}}+1}.
	\label{eq:eps-freq-final}
\end{align}

Combining Eqs.~\ref{eq:Delta-sup-eps}, \ref{eq:eps-weight-final} and
Eq.~\ref{eq:eps-freq-final}, we get the final uniform bound

\begin{equation}
	\sup_{\vec{\theta}} \big|L(\vec{\theta}) - L_{w_{\mathrm{cut}},\delta_{\mathrm{cut}}}(\vec{\theta})\big|\le \frac{C_0	\left(
		A^{w_{\mathrm{cut}}+1}
		+ B^{\delta_{\mathrm{cut}}+1}
		\right)
	}{(1-A)(1-B)}
	\label{eq:truncation-bound-app}
\end{equation}

In other words, under Assumption~\ref{ass:weight-frequency-decay} the truncation
error induced by the cutoffs $w_{\mathrm{cut}}$ and $\delta_{\mathrm{cut}}$ decays
\emph{exponentially} in both cutoffs, up to a pre-factor that depends only on
the system size and the decay constants. This makes precise the heuristic
argument that increasing $w_{\mathrm{cut}}$ and $\delta_{\mathrm{cut}}$ yields
rapidly diminishing error.\\

\subsection{Gradients under Pauli-weight and trigonometric-degree truncation}
\label{sec:gradient_truncation}
The parameter-shift rule, when applied to the symbolically propagated
observable, yields the same gradient of the truncated surrogate
$\tilde{L}_{w_{\mathrm{cut}},\delta_{\mathrm{cut}}}(\vec{\theta})$ (Appendix
\ref{sec:paramshift}), for any gate set admitting a valid parameter-shift rule.
In general, however,
\begin{equation}
	\nabla_{\vec{\theta}} \tilde{L}_{w_{\mathrm{cut}},\delta_{\mathrm{cut}}}(\vec{\theta}) \neq \nabla_{\vec{\theta}} L(\vec{\theta}),
\end{equation}
\newline
\noindent
because the latter still contains contributions from Pauli words with $w(P) >
	w_{\mathrm{cut}}$ or $\delta(P) > \delta_{\mathrm{cut}}$. We now bound the
difference between these two gradients.

Using the Fourier representation of $L(\vec{\theta})$ and its truncated version,
as introduced above, the $j$-th component of the exact gradient can be written
as
\begin{equation}
	\frac{\partial L}{\partial \theta_j}(\vec{\theta})= \sum_{P,k} i k_j\,a_{P,k}\,e^{i k \cdot \vec{\theta}},
	\label{eq:grad_exact_component}
\end{equation}
\noindent
while for the truncated surrogate one obtains
\begin{equation}
	\frac{\partial \tilde{L}_{w_{\mathrm{cut}},\delta_{\mathrm{cut}}}} {\partial \theta_j}(\vec{\theta}) = \!\!\!\! \sum_{\substack{P,k:\\
			w(P) \le w_{\mathrm{cut}},\  \\ \|k\|_1 \le \delta_{\mathrm{cut}}}}\!\!\!\! i k_j\,a_{P,k}\,e^{i k \cdot \vec{\theta}}.
	\label{eq:grad_truncated_component}
\end{equation}
Their difference is therefore
\begin{equation}
	\frac{\partial L}{\partial \theta_j}(\vec{\theta})-\frac{\partial\tilde{L}_{w_{\mathrm{cut}},\delta_{\mathrm{cut}}}}
	{\partial \theta_j}(\vec{\theta}) = \!\!\!\! \sum_{\substack{P,k:\\
			w(P) > w_{\mathrm{cut}} \ \\[.1cm] \text{or}\\[.1cm] \|k\|_1 > \delta_{\mathrm{cut}}}} \!\!\!\! i k_j\,a_{P,k}\,e^{i k \cdot \vec{\theta}}.
	\label{eq:grad_error_sum}
\end{equation}

Taking the supremum over all $\vec{\theta}$ and using $|e^{i k \cdot
	\vec{\theta}}| = 1$ gives
\begin{equation}
	\sup_{\vec{\theta}}\left|\frac{\partial L}{\partial \theta_j}(\vec{\theta})-\frac{\partial \tilde{L}_{w_{\mathrm{cut}},\delta_{\mathrm{cut}}}}
	{\partial \theta_j}(\vec{\theta})\right|\le \sum_{\substack{P,k:\\w(P) > w_{\mathrm{cut}}\\[.1cm] \text{or}\\[.1cm] \|k\|_1 > \delta_{\mathrm{cut}}}}|k_j|\,|a_{P,k}|.
	\label{eq:grad_error_sup}
\end{equation}

Under the locality-induced decay of Fourier coefficients assumed in Sec.
\ref{sec:jamal_truncation}, the right-hand side of Eq.~\ref{eq:grad_error_sup}
is dominated by convergent geometric tails in $w_{\mathrm{cut}}$ and
$\delta_{\mathrm{cut}}$. Consequently, there exist constants $A_j,B_j > 0$ and
\mbox{$0 < \rho_w,\rho_\delta < 1$} such that

\begin{equation}
	\sup_{\vec{\theta}}
	\left|\frac{\partial L}{\partial \theta_j}(\vec{\theta})-\frac{\partial \tilde{L}_{w_{\mathrm{cut}},\delta_{\mathrm{cut}}}}{\partial \theta_j}(\vec{\theta})
	\right|\le A_j\,\rho_w^{\,w_{\mathrm{cut}}+1} + B_j\,\rho_\delta^{\,\delta_{\mathrm{cut}}+1}.
	\label{eq:final_grad_error_bound}
\end{equation}

In particular, the gradient of the double-truncated surrogate converges
uniformly to the gradient of the exact objective as
$w_{\mathrm{cut}},\delta_{\mathrm{cut}} \to \infty$, so that the parameter-shift
rule provides a controllable approximation to $\nabla_{\vec{\theta}} L$ via
$\nabla_{\vec{\theta}} \tilde{L}_{w_{\mathrm{cut}},\delta_{\mathrm{cut}}}$.

\end{document}